\documentclass[conference]{basi}
\usepackage[british]{babel}
\usepackage[varg]{txfonts}
\usepackage{rotating}
\usepackage{dcolumn}
\begin{document}
\title[Back-reactions of dynamo]{Back-reactions of dynamo-generated magnetic fields:
            Torsional oscillations and variations in meridional circulation}
\author[A.~R.~Choudhuri]%
       {Arnab Rai Choudhuri\thanks{email: \texttt{arnab@physics.iisc.ernet.in}}\\
       Department of Physics, Indian Institute of Science, Bangalore--560012, India}

\pubyear{2011}
\volume{00}
\pagerange{\pageref{firstpage}--\pageref{lastpage}}

\date{Received \today}

\maketitle
\label{firstpage}
\def\pa{\partial}
\def\vp{v_{\phi}}
\def\mc{meridional circulation}

\begin{abstract}
The periodically varying Lorentz force of the periodic solar magnetic field
generated by the solar dynamo can induce two kinds of motions: torsional
oscillations and periodic variations in the \mc.  Observational evidence now
exists for both these kinds of motions.  We discuss our ongoing effort in
theoretically studying the variations of the \mc.  Then we present our theoretical
model of torsional oscillations, which addresses the question why these
oscillations start before sunspot cycles at latitudes higher than where
sunspots are seen.
\end{abstract}


\section{Introduction}

The solar cycle is associated with a periodically varying magnetic field
produced by the solar dynamo. This periodically varying magnetic field must
give rise to a periodically varying Lorentz force.  We consider the possible
kinds of motions induced by this periodically varying Lorentz force.

To figure out the kinds of motions induced, we need to look at the Navier--Stokes
equation with the periodically varying Lorentz force inserted in it.  We can
consider two kinds of motions separately: (i) the motions in the $\phi$ direction,
and (ii) the motions in $(r, \theta)$ plane.  It may be mentioned that the equation
of continuity has to be satisfied along with the Navier--Stokes equation.  The
equation of continuity ensures that motions in $r$ and $\theta$ directions are
coupled together.  So we cannot treat those two directions separately.  However,
motions in the $\phi$ direction can be treated separately.

The basic motion in the $\phi$ direction is the differential rotation, which
can be modified by the periodically varying Lorentz force.  Such a periodic
variation of the differential rotation---known as {\em torsional oscillations}---has
been known for about three decades since its discovery at the surface by
Howard \& LaBonte (1980).  On the other hand, the basic motion in the $(r, \theta)$
plane is the meridional circulation and only recently evidence has started
coming that there is a periodic variation of the meridional circulation
with the solar cycle (Hathaway \& Rightmire 2010). Soon after the discovery
of torsional oscillations, some authors started exploring their
theoretical implications (Yoshimura 1981; Sch\"ussler 1981; Tuominen \&
Virtanen 1984).  Due to the rapid developments in dynamo theory in the last
few years, this subject needs revisiting.

We (Karak \& Choudhuri) are working on a theoretical model of the variation
of meridional circulation due to the Lorentz force.  Some of the basic concerns are
discussed in \S2.  Then in \S3 we summarize our recent work on torsional
oscillations (Chakraborty, Choudhuri \& Chatterjee 2009a).

\section{Variations of meridional circulation}

The meridional circulation of the Sun plays a crucial role in the flux transport
dynamo.  However, its theory is still rather poorly understood. 
Helioseismology provides some information about the meridional circulation 
in the upper layers of the convection zone (Giles et al. 1997; Braun \& Fan 1998; 
Gonzalez Hermandez et al. 2006; Svanda, Kosovichev \& Zhao 2007).  But there
is no reliable observational data available yet on the nature of the meridional
circulation at the bottom of the convection zone, which is crucial for the
dynamo (Nandy \& Choudhuri 2002).

The classic investigation by Kitchatinov \& R\"udiger (1995) showed that
the meridional circulation arises out of a slight imbalance between two large
terms, requiring a pole-equator temperature difference of about $5^{\circ}$.
A consequence of this is that even small fluctuations in any of these large
terms may cause significant variations in meridional circulation. On the basis
of indirect evidence, Karak (2010) concluded that there have been large variations
in meridional circulation in the last few centuries.  The variations in
meridional circulation also seem to be the main reason behind the well-known
Waldmeier effect of solar cycles (Karak \& Choudhuri 2011a). 

Apart from these random variations of meridional circulation, we expect some
systematic periodic variations with the solar cycle.  Only recently Hathaway
\& Rightmire (2010) and Basu \& Antia (2010) have presented evidence for this.
It appears that the meridional circulation near the surface becomes somewhat
weaker at the time of the sunspot maximum.  We expect a strong toroidal field
at the bottom of the convection zone at the time of the sunspot maximum.  Such
a toroidal field has a Lorentz force directed towards the rotation axis and
has a tendency of slipping poleward (van Ballegooijen \& Choudhuri 1988).
This poleward slipping tendency will oppose the equatorward meridional
circulation at the bottom of the convection zone.  We (Karak \& Choudhuri)
are now carrying out a detailed calculation to investigate whether the surface
observations of the reduction of meridional circulation at the time of sunspot
maximum can be explained as arising out of this opposition to meridional
circulation due to the poleward slip tendency of the strong toroidal field
at the bottom of the convection zone.

It may be mentioned that Nandy, Mu\~noz-Jaramillo \& Martens (2011) 
have assumed in a recent work 
that the \mc\ changes randomly at every solar maximum.  We disagree
with this assumption and believe that the \mc\ decreases at the solar
maximum due to the Lorentz force of the magnetic fields in a systematic
deterministic way. If the Lorentz force quenches the \mc\ at the time
of the solar maximum, one important question is whether this will have
any effect on the dynamo.  Recently Karak \& Choudhuri (2011b) performed
a dynamo simulation by assuming that the amplitude of the \mc\ is
quenched according to the equation
\begin{equation}
v_0 =  v_0^\prime/[1+(\overline{B} /B_0)^2],
\end{equation}
where $\overline{B}$ is the average value of the toroidal field at
the bottom of the convection zone.  If the value of the turbulent diffusivity
is assumed to be sufficiently small ($\eta \sim 10^{10}$--$10^{11}$ cm$^2$ s$^{-1}$), 
then the dynamo becomes unstable on including such a quenching of the \mc.
The dynamo remains stable only if the turbulent diffusivity is in the
range $\eta \sim 10^{12}$--$10^{13}$ cm$^2$ s$^{-1}$.  This provides
another argument in favour of a high-diffusivity dynamo, strengthening a
case already made by several authors (Jiang, Chatterjee \& Choudhuri 2007;
Goel \& Choudhuri 2009; Hotta \& Yokoyama 2010; Karak 2010).

\section{Torsional oscillations}

The small periodic variation in the Sun's rotation with the
sunspot cycle, first discovered on the solar surface by Howard
\& LaBonte (1980), is called torsional oscillations.
Helioseismology has now established its existence throughout
the convection zone (see Howe et al.\ 2005 and references
therein). Its amplitude
near the surface is of order 5 m s$^{-1}$ or about 1\% of
the angular velocity. Apart from
the equatorward-propa\-gating branch which moves with the sunspot
belt, there is also a 
poleward-propagating branch at high latitudes. One intriguing
aspect of the equatorward-propagating branch is that it
begins a couple of years before the sunspots of a particular
cycle appear and at a latitude higher than where
the first sunspots are seen.  The top panel of Fig.~1 shows
the torsional oscillations at the solar
surface with the butterfly
diagram of sunspots. If the torsional oscillation
is caused by the Lorentz force of the dynamo-generated
magnetic field as generally believed, 
then the early initiation of this oscillation
at a higher latitude does 
look like a violation of causality! The main aim 
of our recent work (Chakraborty, Choudhuri \& Chatterjee 2009a, 2009b)
has been to explain
this which could not be explained by the earlier theoretical
models (Durney 2000; Covas et al.\ 2000; Bushby 2006; Rempel
2006). 

\begin{figure}
\center
\includegraphics[width=10cm]{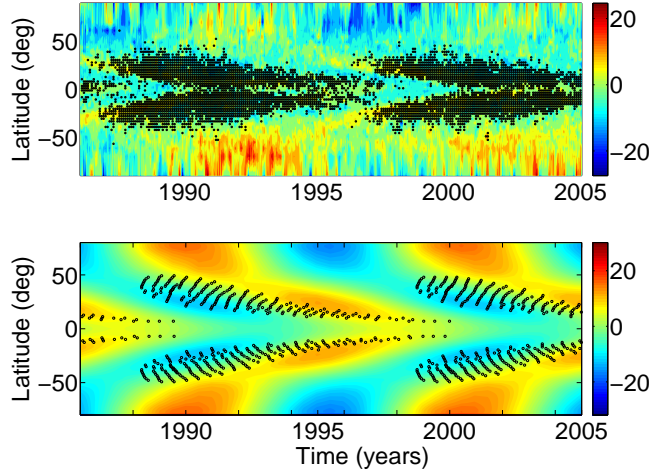}
\caption{The time-latitude plot of torsional oscillations on the
solar surface with the butterfly diagram of sunspots superposed
on it. The upper panel is based on observational data of surface
velocity $\vp$ measured at Mount Wilson Observatory (courtesy:
Roger Ulrich).  The bottom panel is from our theoretical simulation.}
\end{figure}
 
Our calculations are based on the dynamo model presented
by Chatterjee, Nandy \& Choudhuri (2004).  
In order to model torsional oscillations, in addition to the
basic equations of the dynamo, we simultaneously have to solve
the $\phi$ component of the Navier--Stokes equation in the form
$$\rho \left\{ \frac{\pa \vp}{\pa t} + D_v [\vp] \right\} = 
D_{\nu} [\vp] + ({\bf F}_L)_{\phi},\eqno(2)$$ 
where 
$D_v [\vp]$ is the term corresponding to advection by the meridional circulation,
$D_{\nu} [\vp]$ is the diffusion term, and 
$({\bf F}_L)_{\phi}$ is the $\phi$ component of the Lorentz force.
\def\Bp{B_{\phi}}
If the magnetic field is assumed
to have the standard form
$${\bf B} = B (r, \theta, t){\bf e}_{\phi} + \nabla \times 
[A(r, \theta, t){\bf e}_{\phi}], \eqno(3) $$
then the Lorentz force is given by the Jacobian
$$4 \pi ({\bf F}_L)_{\phi} = \frac{1}{s^3} J \left( \frac{s B_{\phi}, 
s A }{r, \theta} \right), \eqno(4)$$
where $s= r \sin \theta$. On the basis of flux tube simulations
suggesting that the magnetic field in the tachocline should be
of order $10^5$ G (Choudhuri \& Gilman 1987; Choudhuri 1989;
D'Silva \& Choudhuri 1993), it is argued by Choudhuri (2003)
that the magnetic field has to be
intermittent in the tachocline. Hence the full expression
of the Lorentz force involves a filling factor as explained
by Chakraborty, Choudhuri \& Chatterjee (2009a).

\begin{figure}
\begin{minipage}[b]{0.5\textwidth}
\includegraphics[height=4cm,width=6cm]{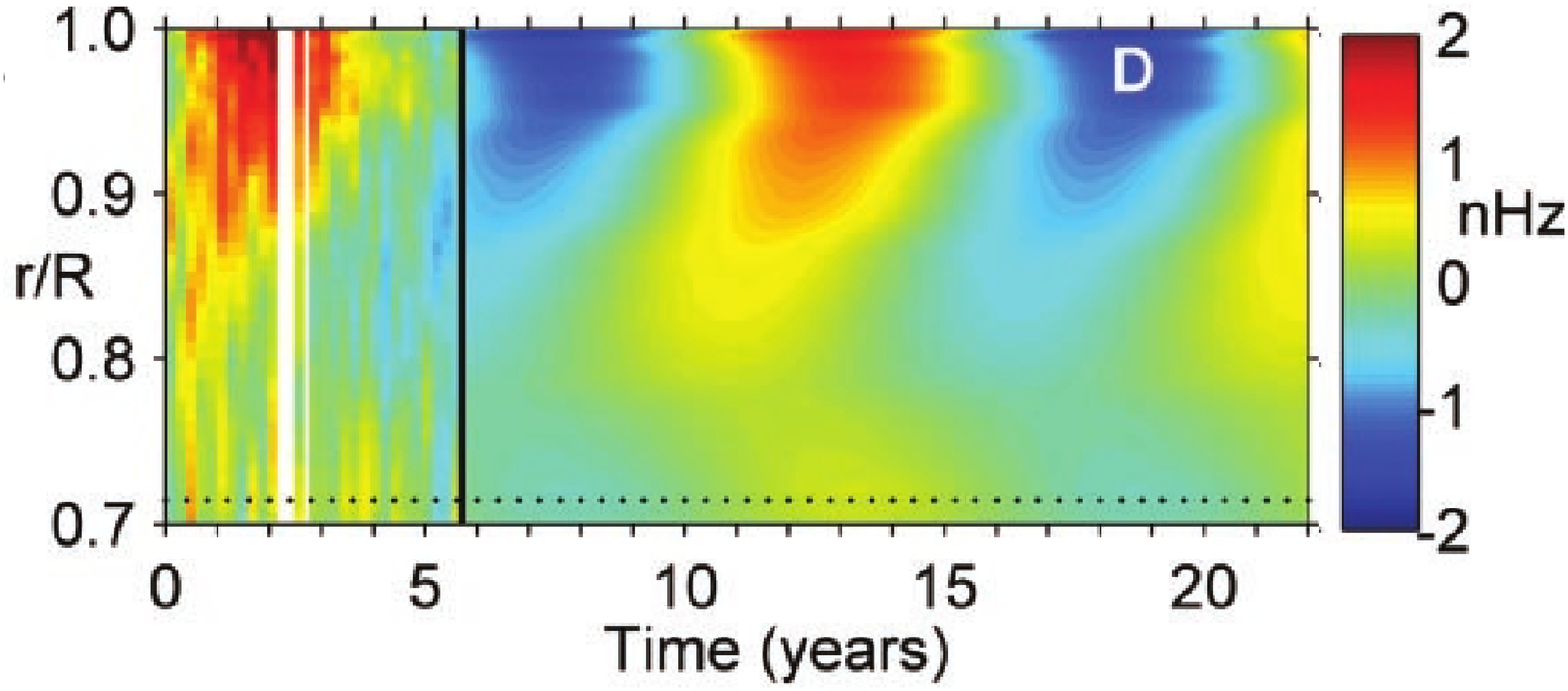}
\end{minipage}
\begin{minipage}[b]{0.5\textwidth}
\includegraphics[height=4cm,width=6cm]{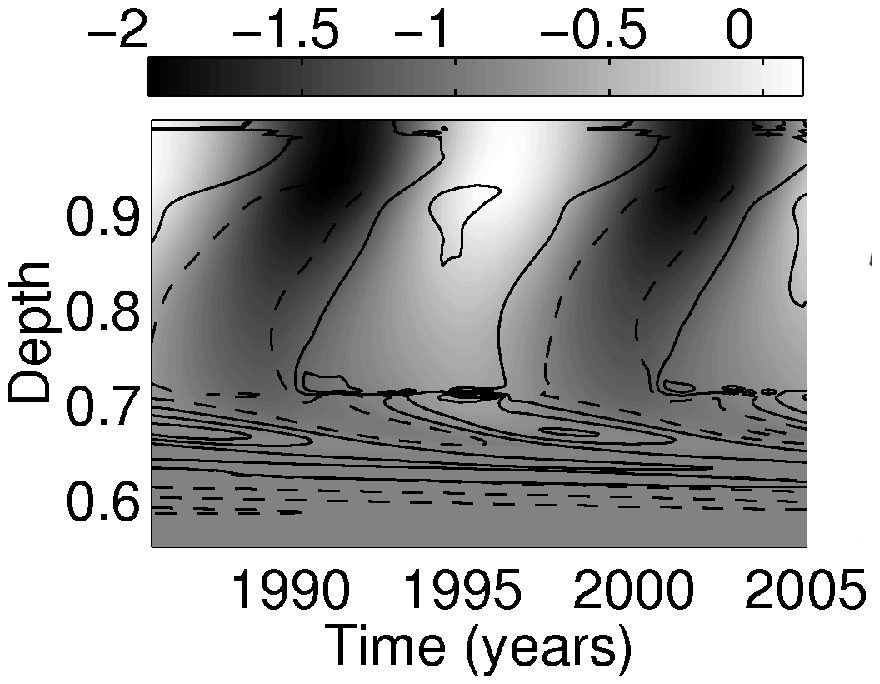}
\end{minipage}
\caption{The depth-time plot of torsional oscillations
at latitude $20^{\circ}$.
The left panel from Vorontsov et al.\ (2002)  
is based on SOHO observations,
whereas the right panel from Chakraborty, Choudhuri \& Chatterjee (2009b) is based
on our theoretical simulation.
The solid and dashed lines in the right panel indicate the Lorentz force
(positive and negative values respectively).}
\end{figure}

Our theoretical model incorporates a hypothesis proposed by
Nandy \& Choudhuri (2002), which is essential for explaining the
early initiation of the torsional oscillation at high latitudes.
According to this Nandy--Choudhuri (NC) hypothesis, the
meridional flow penetrates into the stable layers below the convection 
zone at high latitudes.  This causes the formation of toroidal 
field in the high-latitude tachocline. 
Sunspots form a few years later when this field is advected to 
lower latitudes and brought inside the convection zone. We also
assume that the stress of the magnetic field formed in the tachocline
is carried upward by Alfv\'en waves propagating along vertical flux
concentrations conjectured by Choudhuri (2003).

The incorporation of the NC hypothesis in our theoretical model
causes magnetic stresses to build up at higher latitudes before 
sunspots of the cycle appear, leading to the early 
initiation of torsional oscillations.
The bottom panel of Fig.~1 shows theoretical results of torsional
oscillations at the surface with the theoretical butterfly diagram.
This bottom panel can be compared with the observational upper
panel in Fig.~1. Our theoretical model also gives a satisfactory
account of the evolution of torsional oscillations within the convection
zone.  The depth-time plot of torsional oscillations at a certain
latitude given in Fig.~3 of Chakraborty, Choudhuri \& Chatterjee (2009b) compares favourably with the 
observational plot given in Fig.~3(D) of Vorontsov et al.\ (2002). 
This is reproduced here in Fig.~2 for completeness.  We end by pointing
out that our theoretical model of torsional oscillations attempts to
explain various aspects of observational data in much greater detail
than what had been attempted in previous theoretical models.

\end{document}